\documentclass[12pt]{article}
\pdfoutput=1
\usepackage{graphicx}
\usepackage{epsfig,cite}
\usepackage{amssymb}
\usepackage{amsmath}
\usepackage{dsfont}
\usepackage{multirow}
\usepackage{color}
\usepackage{subfigure,amstext,alltt,setspace}
\usepackage{amsbsy}
\usepackage{comment}
\usepackage{fullpage}
\usepackage{array}
\usepackage{booktabs,multirow,tabularx}
\usepackage{hyperref}
\usepackage{slashed}
\usepackage{url}

\newcommand{\order}[1]{{\cal O}(#1)}

\allowdisplaybreaks[1]
\begin{document}

\begin{flushright}
  TIF-UNIMI-2018-3\\
  DAMTP-2018-13
\end{flushright}

\vspace*{.2cm}

\begin{center}
  {\Large \bf{$Z$ boson production in bottom-quark fusion:\\
      a study of $b$-mass effects beyond leading order}}
\end{center}

\vspace*{.7cm}

\begin{center}
  Stefano Forte$^{1}$, Davide Napoletano$^2$ and Maria Ubiali$^{3}$
  \vspace*{.2cm}

  \noindent
      {\it
        $^1$ Tif Lab, Dipartimento di Fisica, Universit\`a di Milano and\\ 
        INFN, Sezione di Milano,
        Via Celoria 16, I-20133 Milano, Italy\\
        $^2$ IPhT, CEA Saclay, CNRS UMR 3681,\\ F-91191, Gif-Sur-Yvette, France\\
        $^3$ DAMTP, University of Cambridge,\\ Wilberforce Road, Cambridge, CB3 0WA, UK\\}

      \vspace*{3cm}

      %\begin{center}
      {\bf Abstract}
\end{center}

\noindent
We compute the total cross-section for $Z$ boson production in
bottom-quark fusion, applying to this case the method we previously
used for Higgs production in bottom fusion. Namely, we  match, through
the FONLL procedure,
the  next-to-next-to-leading-log five-flavor
scheme result, in which  the $b$~quark is
treated as a massless parton, with the next-to-leading-order
$\order{\alpha_s^3}$  
four-flavor scheme computation in which bottom is treated as a massive
final-state particle.
Also, we add to our formalism the possibility of varying the
heavy quark matching scale. 
The results obtained with the FONLL formalism can thus be compared
directly
 to recent results obtained in various approximations, and used as a
proxy 
to assess and discuss the issues of scale dependence and treatment of heavy
quarks. We also use our results in order 
to improve the prediction for the total
$Z$ production cross-section.

\pagebreak

%\tableofcontents

The production of a $Z$ boson is one of the main standard candles at the LHC,
and is now measured at the sub-percent level. The main production mode
is through quark-anti-quark fusion, of which the bottom-initiated contribution
accounts to ${\cal O}(4\%)$ of the total cross section.
This is  a small but non negligible fraction of the total cross
section, and
its contribution affects both the normalisation and the shape
of the kinematic distributions.  Therefore a precise estimate of the
bottom-initiated contribution is important for precision physics, for
example in the determination of the $W$ mass~\cite{Bagnaschi:2018dnh}.
This process is thus an ideal test case for matched computations,
recently applied to Higgs production 
in bottom quark
fusion~\cite{Forte:2015hba,Forte:2016sja,Bonvini:2015pxa,Bonvini:2016fgf}. As
we shall show here, it  provides a theoretically transparent setting
for the discussion of issues of  choice of scheme and scale in  the
treatment of heavy quark contributions.

Like any process involving bottom quarks at the matrix-element level, 
the bottom-initiated $Z$ production process 
may be computed using two different  factorisation schemes, which we
refer to, as usual, as four- and five-flavour schemes for short. In the
four-flavour scheme (4FS), the $b$~quark is treated as a massive
object, which 
decouples from QCD perturbative evolution. Calculations in this scheme
are thus performed by only including
the four lightest flavour together with the gluon  in evolution
equations for parton distributions (PDFs), and in the running of
$\alpha_s$,
so $n_f=4$ in the QCD $\beta$ function.
In the five-flavour scheme (5FS), instead, the $b$~ quark is treated on
the same footing as other light quark flavours, there is a $b$~PDF, and
$n_f=5$ in evolution equations for PDFs and in the QCD $\beta$ function.

In matched calculations, both scheme are combined, in such a way that
the result differs by that of each of the two schemes by terms which
are sub-leading with respect to the accuracy of either of them. 
The FONLL scheme,
first proposed for  heavy quark production in hadronic
collisions~\cite{Cacciari:1998it} has the advantage of being
universally applicable; also, it allows for the matching of
four- and five-flavour computations performed at any combination of
individual perturbative orders. It has been extended to deep-inelastic
scattering in Ref.~\cite{Forte:2010ta} (also
including~\cite{Ball:2015tna,Ball:2015dpa} the case in which the heavy
quark PDF is independently parametrised) and, as mentioned, it has
been
used in
Refs.~\cite{Forte:2015hba,Forte:2016sja} for the computation of the
total cross-section for Higgs production in bottom
quark fusion.

\begin{figure}
  \begin{center}
    \includegraphics[width=0.35\textwidth]{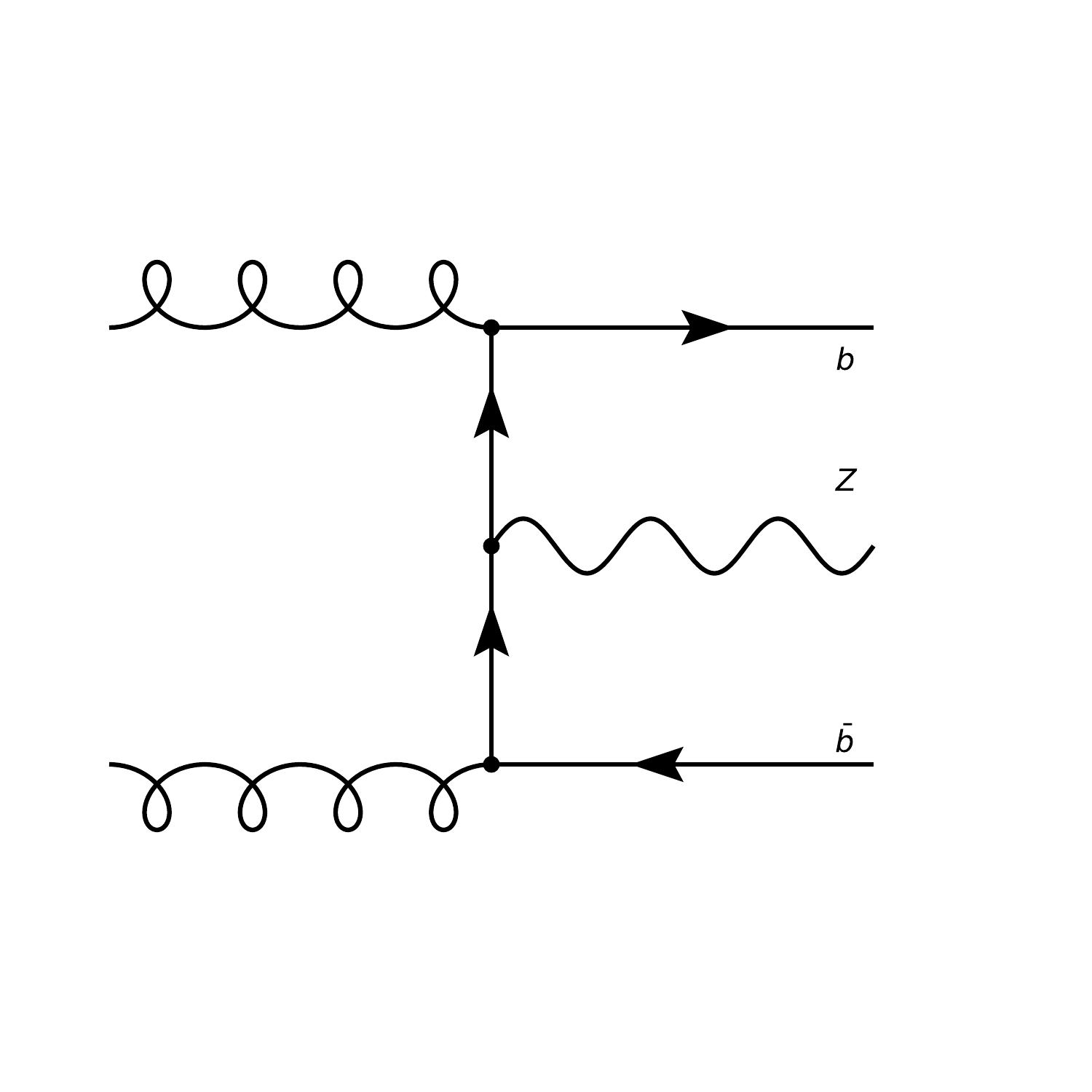}
    \includegraphics[width=0.35\textwidth]{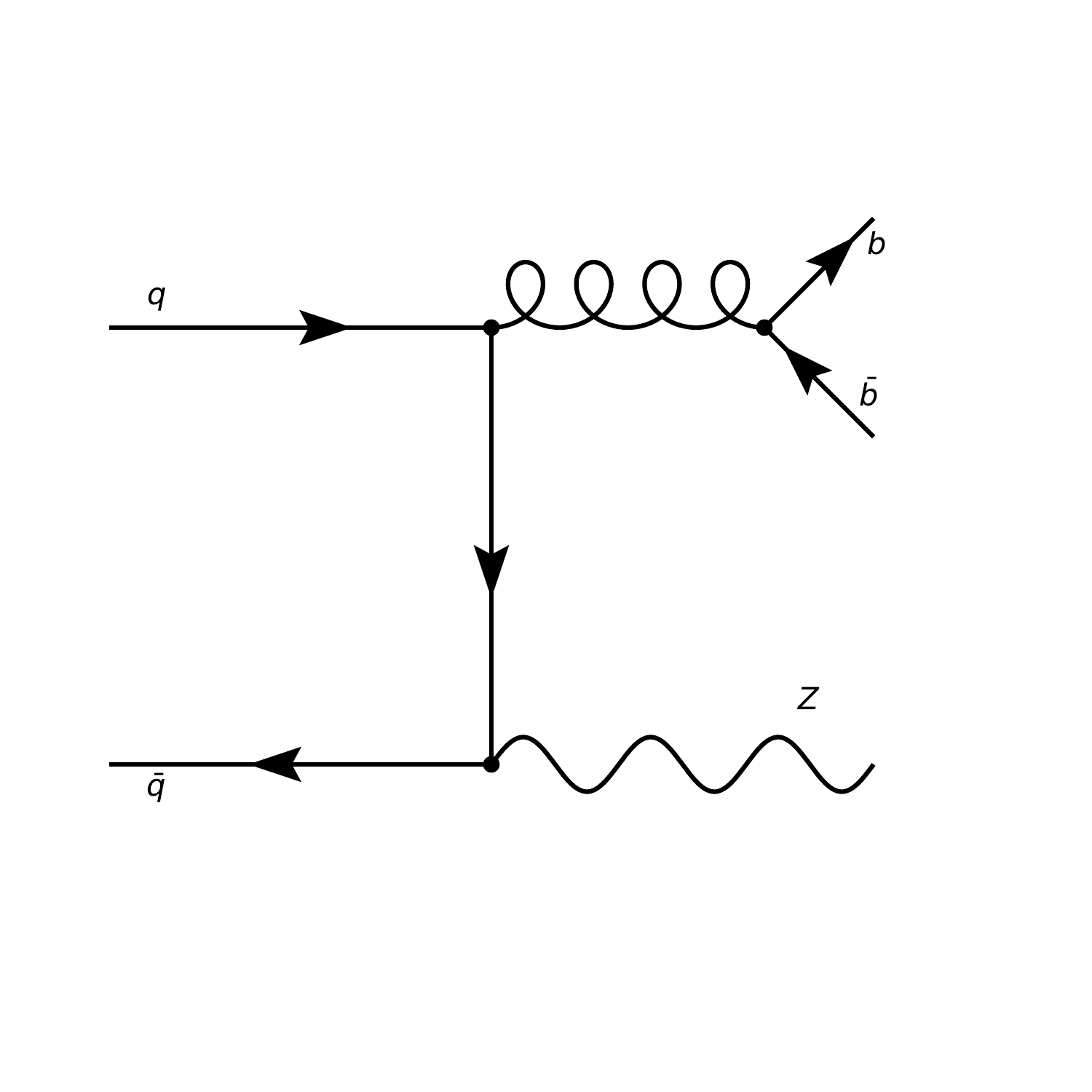} 
    \caption{\label{fig:fusvsfs} Leading-order diagrams for $Zb\bar b$ 
 production in the four-flavor scheme: bottom fusion (left)
 and production from the initail state light quark (right).
    }
  \end{center}
\end{figure}
Here, the methodology of
Refs.~\cite{Forte:2015hba,Forte:2016sja} is applied to $Z$
production. When comparing Higgs to $Z$ production in bottom quark
fusion some care must be taken in defining exactly which process is
being considered. 
Indeed, in the case of Higgs production the bottom fusion
cross-section can be equivalently viewed as the cross section for
associate production in conjunction with a pair of $b$ jets, i.e. as the
$H b\bar b$ cross section. In the case of $Z$ production in the
four-flavor scheme, on top of the leading-order $Z$ production diagram
in bottom fusion  there is also a process in the quark-antiquark
channel which produces the same $Z b\bar b$ final state, but in which
the $Z$ is radiated by  initial-state  light quarks and there are no
$b$ quarks in the initial state (see Fig.~\ref{fig:fusvsfs}).

Hence, for $Z$ production, unlike Higgs production, one may consider,
at least in principle, two distinct processes. The first is $Z$ production
in bottom fusion, defined as $Z$ production in the case in
which the coupling of the $Z$ to all quarks but bottom vanishes. In
this case, only the diagram shown on the left of
Fig.~\ref{fig:fusvsfs} contributes (as for Higgs production). The
second is $Z\bar b$ production, defined as the process with a $Z$ and a
bottom quark-antiquak pair in the final state, in which case both
diagrams in Fig.~\ref{fig:fusvsfs} contribute. In the sequel we will
consistently refer to the first definition (the one which is similar
to Higgs) as $Z$ production in bottom fusion, and to the second (the
definition based on the final state) as $Zb\bar b$ production.

The possibility of
separating experimentally the light quark- and gluon-initiated
contributions of Fig.~\ref{fig:fusvsfs} to $Zb\bar b$ production
has been discussed in
Ref.~\cite{Chatrchyan:2013zja}, where it was shown that by choosing
suitable kinematic variables it is possible to select regions in
which the light quark contribution is dominant. However, from a theoretical
point of view, the  $Zb\bar b$ process is problematic because it is not
infrared and collinear safe if the bottom mass is neglected, and thus
it is beset by mass singularities in the 5FS. This is due to the fact
that diagrams in which the bottom quark appears in the final state are
counted as contributions to the  $Zb\bar b$ process, but virtual corrections in
which the $b$ quark circulates in loops but is absent from  the final state
are not, and thus the cancellation of infrared singularities is
incomplete. In the 4FS this leads to mass-singular contributions
which are finite, but enhanced by double logs of the heavy quark
mass. The problem is completely analogous to one which arises when
definig heavy-quark deep inelastic structure functions, and was
discussed in that context in Ref.~\cite{Forte:2010ta}, to which the
reader is referred for a discussion of the way these double logs can
be resummed. 

Here, we will first focus on the construction of a matched computation
of the process of $Z$ production in bottom fusion, closely following
the related case of Higgs production of
Refs.~\cite{Forte:2015hba,Forte:2016sja}. This will provide us with an
interesting case study for issues related to scale choice and the
relevance of matched computations. 
We will then turn to the use of this result as a means to improve the
total $Z$ cross-section, and in particular revisit the issue of the
appropriate inclusion of light-quark initiated contributions to the
$Zb\bar b$ process.

In the 5FS, the $Z$-production cross section has been known up to
next-to-next-to leading order
(NNLO) (i.e. $\order{\alpha_s^2}$) for almost three decades~\cite{Hamberg:1990np} and 
the heavy-quark initiated contribution has been specifically discussed
in 
several papers~\cite{Rijken:1995gi,Stelzer:1997ns,Maltoni:2005wd}.
The next-to-leading order (NLO) ($\order{\alpha_s^3}$) four-flavour scheme
$Zb\bar{b}$ production cross section was originally computed 
in Ref.~\cite{Campbell:2000bg} 
for exclusive 2-jet final states, neglecting the $b$~quark mass. 
The $b$-quark mass was subsequently fully included in
Refs.~\cite{FebresCordero:2008ci,Cordero:2009kv}.

Our first task is thus to use these two results in order to produce a
matched computation for $Z$ production in bottom fusion,
following the procedure we
presented in Refs.~\cite{Forte:2015hba,Forte:2016sja} for the closely
related case of Higgs production. Indeed, the counting of perturbative
orders for these two processes is the same, and many of the Feynman
diagrams are identical, with the only replacement of Higgs Yukawa
couplings with gauge couplings. Following the nomenclature introduced
in Ref.~\cite{Forte:2015hba,Forte:2016sja}  (and originally in
Ref.~\cite{Forte:2010ta} for DIS) we have constructed an FONLL-A
result, which combines the NNLO 5FS with the LO  $\order{\alpha_s^2}$
4FS fully massive computation,  and an FONLL-B, where instead the NNLO
5FS is matched to the full  NLO $\order{\alpha_s^3}$ massive
results. 

Our construction is essentially identical to that of
Refs.~\cite{Forte:2015hba,Forte:2016sja}, to which we refer for
details: it can be obtained from it by simply replacing the matrix elements
for Higgs production with those for gauge boson
production. Specifically, we have computed the 5FS NNLO 
cross-section using the 
code of Ref.~\cite{Maltoni:2005wd}, which we cross-checked both 
at LO and NLO against MG5\_aMC@NLO~\cite{Alwall:2014hca}. For the
massive 4FS LO and NLO we have also used  MG5\_aMC@NLO.
The construction of the FONLL matched results requires the computation
of the massless limit of the massive result:  we have implemented this
in the  public code~\cite{code} used in~\cite{Forte:2016sja}, in an
updated version soon to be made public.
All predictions are obtained using the NNLO NNPDF3.1 PDF 
set~\cite{Ball:2017nwa}. Lastly, PDF with varied thresholds
are obtained using the APFEL evolution library~\cite{Bertone:2013vaa}.
In order to be consistent with the PDF set used we take, in the 4F scheme,
the $b$ pole mass to be $m_b=4.92$~GeV, while the strong coupling
is run at NNLO, with $\alpha_s(m_Z) = 0.118$.

%% edit
A new feature  in comparison to
Refs.~\cite{Forte:2015hba,Forte:2016sja},
is that we have now extended the formalism to allow for variation of
the scale $\mu_b$ at which the 4FS and 5FS schemes are matched.
In contrast,  in previous FONLL implementations, scale was fixed at
the  bottom mass:  $\mu_b=m_b$.
The reason why results depend on a matching scale is that in 
the 5FS the $b$~PDF is not independently
parametrised. Rather, it is assumed that it is radiatively generated
by the gluon. The matching scale is then the scale at which the $b$~PDF
is determined from the gluon. 
The interest in this is twofold. First,  it allows us
to perform  a direct comparison with recent work~\cite{Bertone:2017djs},
in which  he impact of varying the matching scale  is studied in the
5FS,  and in particular it is argued  that it might be
advantagous to choose a very large value $\mu_b\gg m_b$. 
Second, studying the effect  of matching scale variation provides us 
with another handle on the relative size of various contributions to
the matched calculation, which we will study explicitly.
%% arxiv
%% New in comparison to Refs.~\cite{Forte:2015hba,Forte:2016sja}, we have
%% now implemented the possibility of varying the scale $\mu_b$ at which
%% the 4FS and 5FS schemes are matched. This scale was taken to coincide
%% with the bottom mass  $\mu_b=m_b$ in previous FONLL implementations,
%% but there is no fundamental reason for this choice. The reason why
%% results depend on a matching scale is that in 
%% the 5FS the $b$~PDF is not independently
%% parametrised. Rather, it is assumed that it is radiatively generated
%% by the gluon. The matching scale is then the scale at which the $b$~PDF
%% is determined from the gluon. 

The matching condition itself depends on
the matching scale in such a way that, at any given order,  results
are  independent of it up to sub-leading corrections. This dependence
persists in the FONLL matched results, but it is alleviated if the
scale of the process is  not too far from the bottom production
threshold, because then
the FONLL results almost reduces to the exact mass-dependent result in
which the physical threshold is implemented exactly (as shown
expicitly e.g. in Ref.~\cite{Forte:2010ta}). It reappears when
the scale of the process is 
high enough, in which case  the FONLL result reduces to the 5FS, and it
only goes away when computing the matching condition to increasingly
high perturbative order, or by independently parametrising the heavy
quark PDF (indeed, this is the main motivation for independently
parametrising charm~\cite{Ball:2016neh,Ball:2017nwa}).

The generalisation of the FONLL matching formulae of
Refs.~\cite{Forte:2015hba,Forte:2016sja} for a generic choice of
matching scale is given in the Appendix.
Dependence on this
matching scale for Higgs in bottom fusion  was studied explicitly in
Ref.~\cite{Bonvini:2015pxa,Bonvini:2016fgf}. The matching scheme of 
Ref.~\cite{Bonvini:2015pxa,Bonvini:2016fgf}, based on a EFT approach,
was benchmarked in Ref.~\cite{deFlorian:2016spz} to that of
Refs.~\cite{Forte:2015hba,Forte:2016sja} and found to agree with it at
the percent level, hence a very similar dependence is expected for
FONLL.

\begin{figure}
  \begin{center}
    \includegraphics[width=0.7\textwidth]{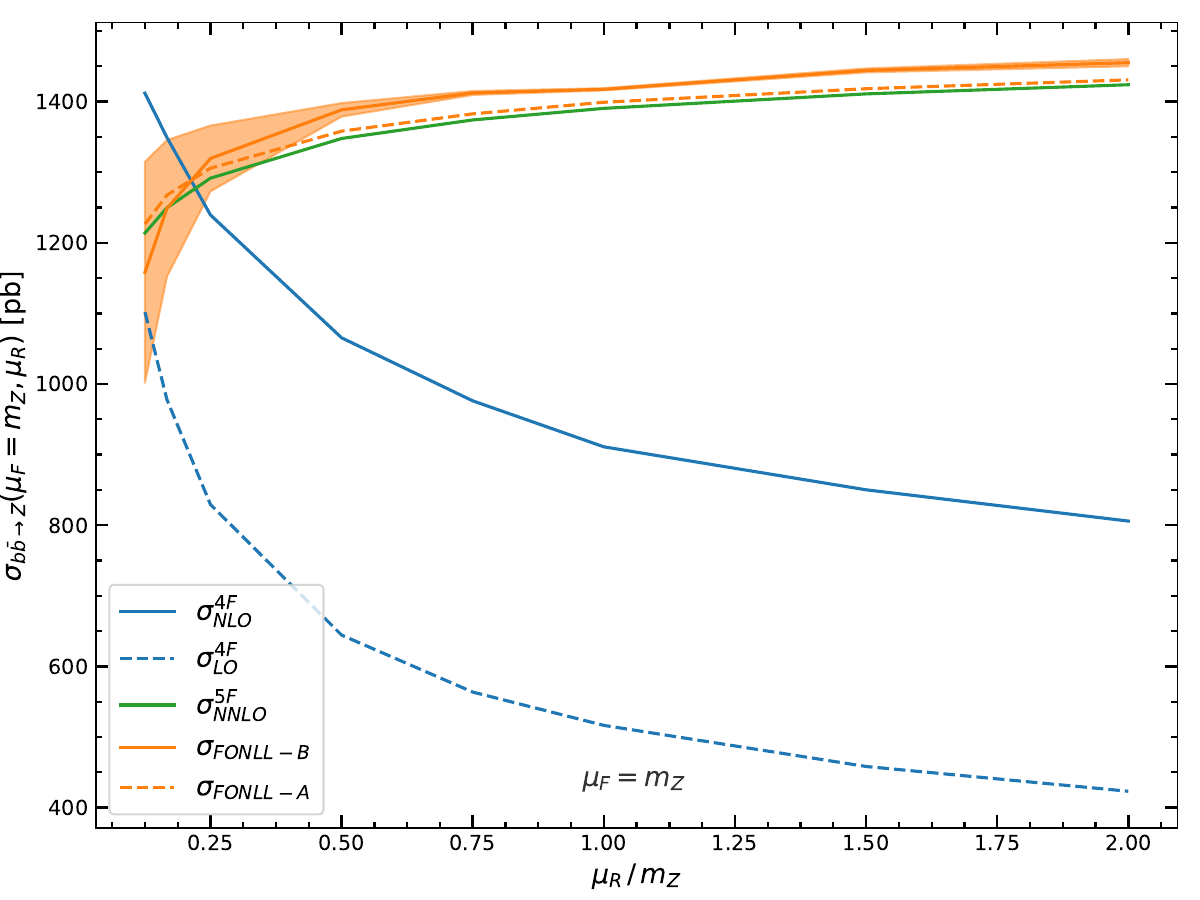}
    \includegraphics[width=0.7\textwidth]{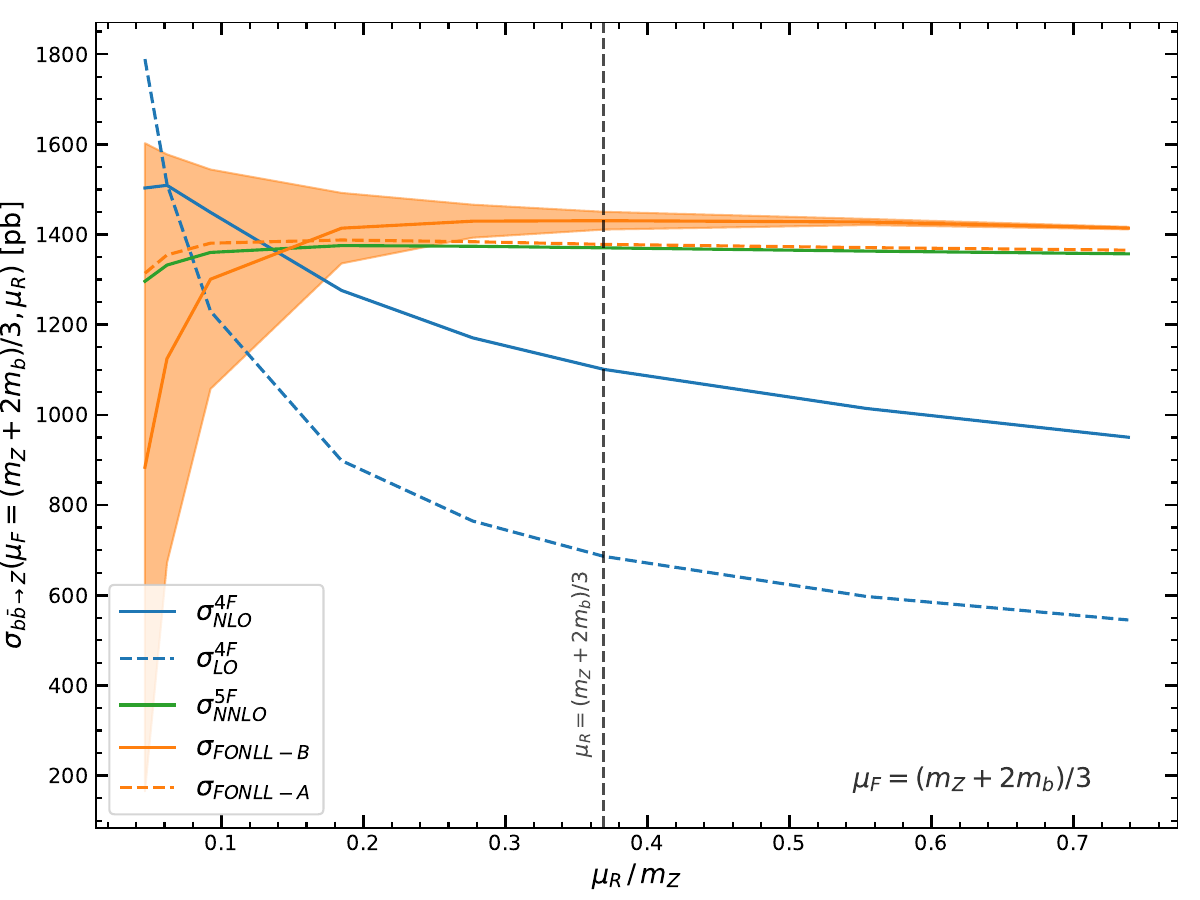} 
    \caption{\label{fig:muR_var} Comparison of the FONLL-A and FONLL-B 
      matched results to each other, to the 4FS LO ($\order{\alpha_s^2}$)
      and NLO  ($\order{\alpha_s^3}$), and to the 5FS NNLO. Results are shown as a function of the
      renormalisation scale, with the factorisation scale fixed at a high value 
      $\mu_F=m_Z$  (top) or a low value  $\mu_F=\frac{(m_Z+2m_b)}{3}$
      (bottom). The band about the FONLL-B result is obtained from two
      different implementations of NLO scale variation that differ by NNLO
      terms (see text) and is thus an estimate on the ambiguity of the
      scale variation itself.
    }
  \end{center}
\end{figure}

\begin{figure}
  \begin{center}
    \includegraphics[width=0.7\textwidth]{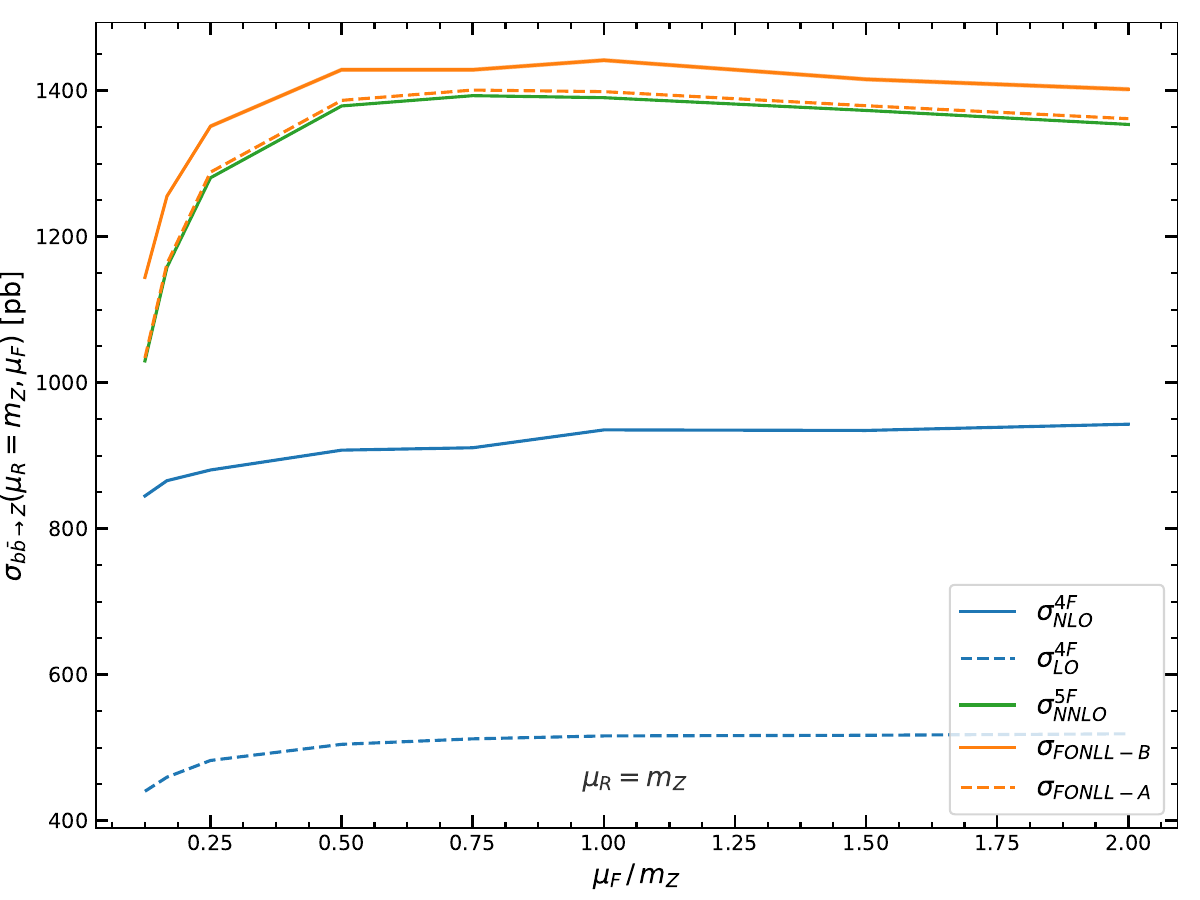}
    \includegraphics[width=0.7\textwidth]{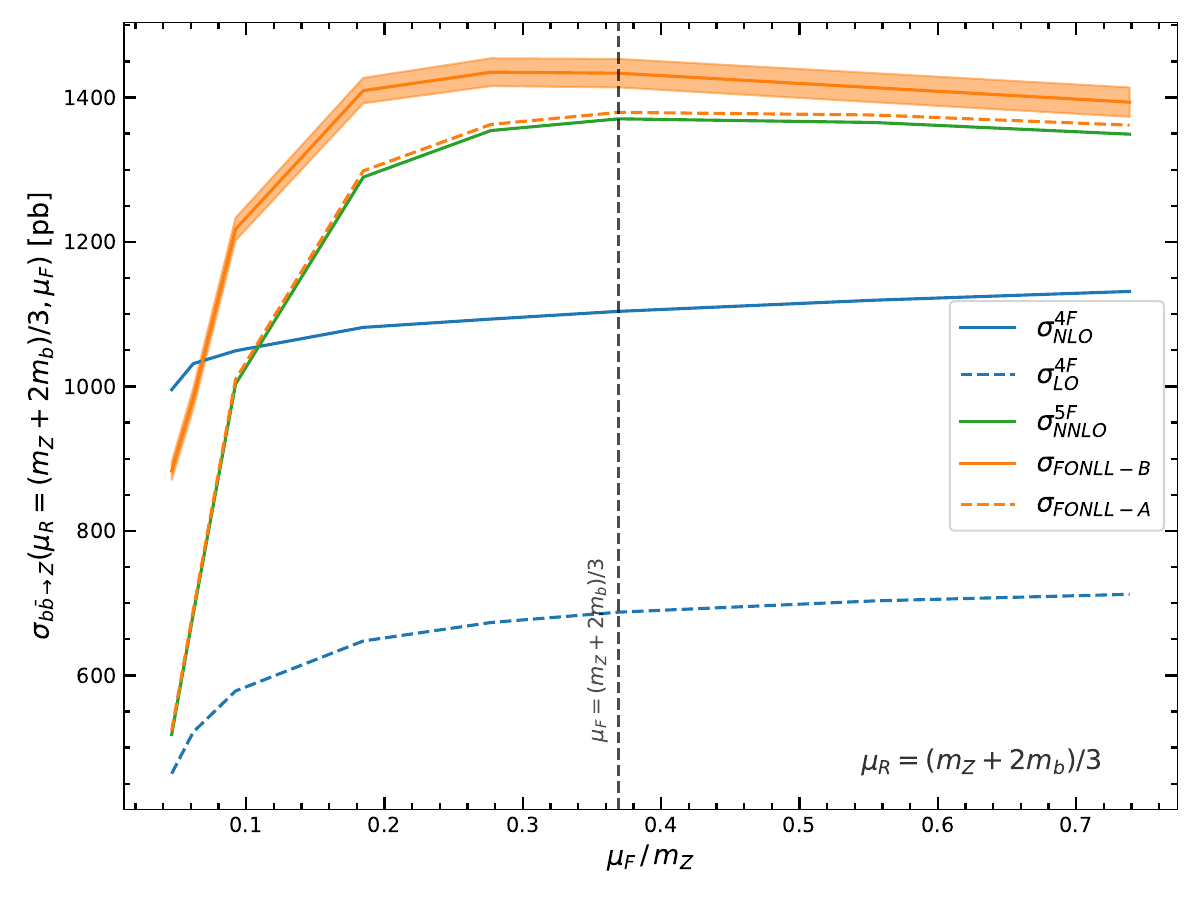} 
    \caption{\label{fig:muF_var} Same as Fig.~\ref{fig:muR_var}, but now
      with the factorisation scale varied with the renormalisation scale
      kept fixed at a high value 
      $\mu_R=m_Z$  (top) or a low value  $\mu_R=\frac{(m_Z+2m_b)}{3}$ (bottom) .}
  \end{center}
\end{figure}

\begin{figure}
  \begin{center}
    \includegraphics[width=0.7\textwidth]{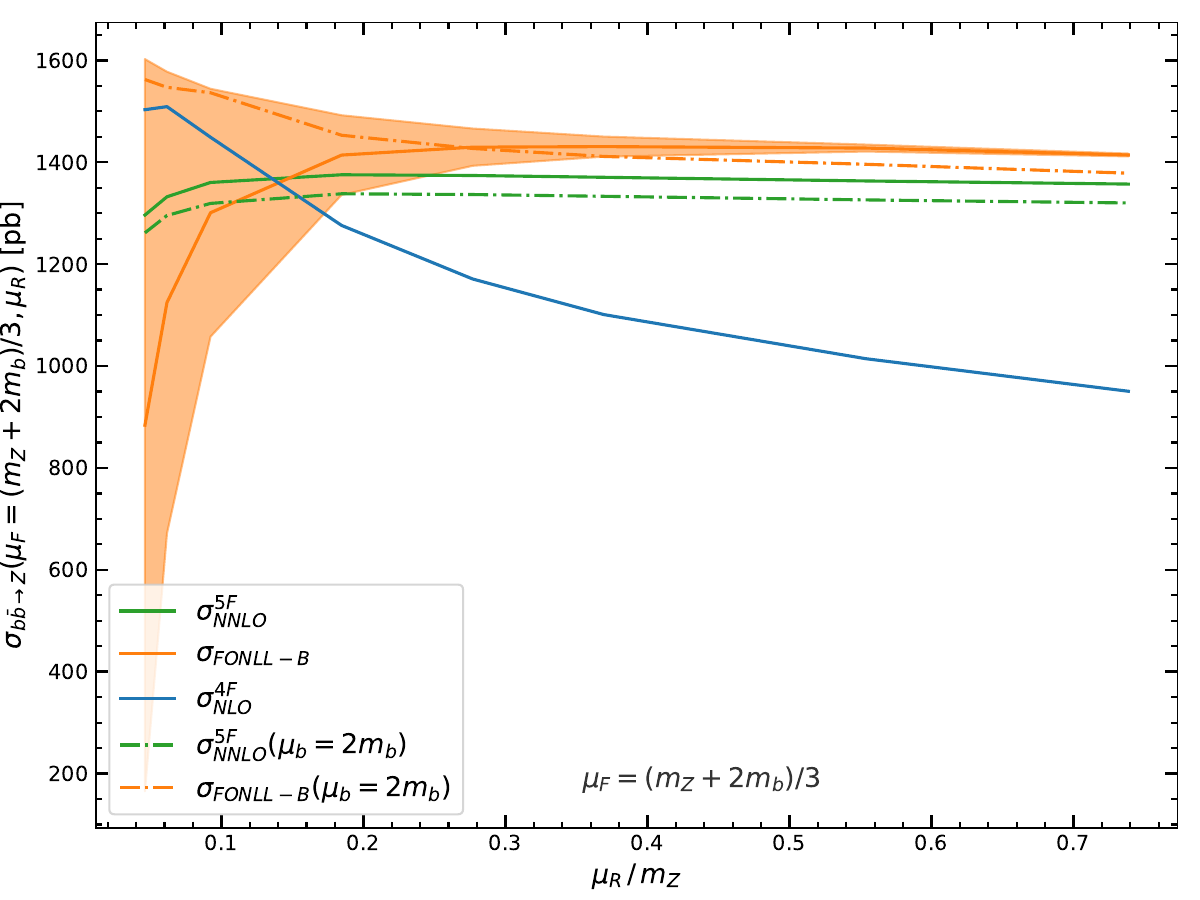}
    \includegraphics[width=0.7\textwidth]{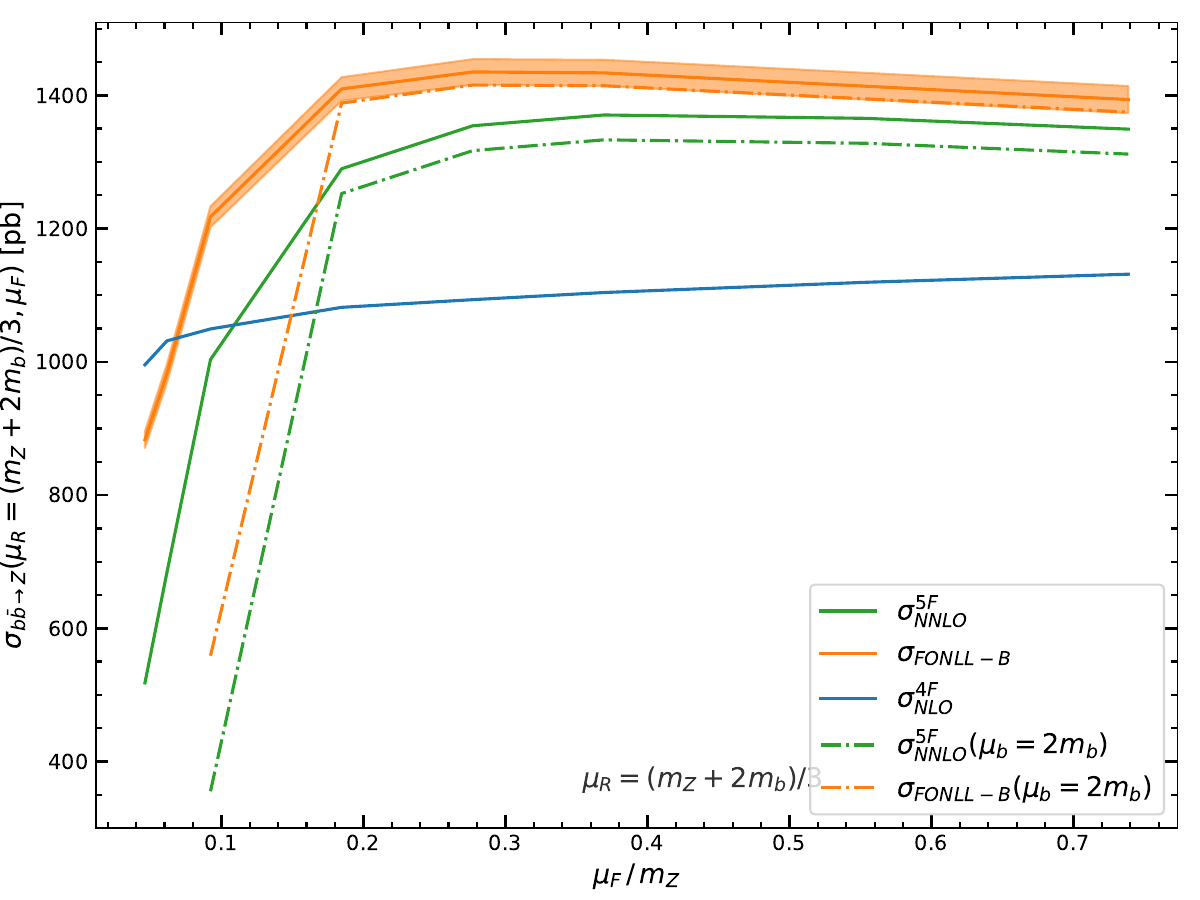} 
    \caption{\label{fig:m_mu_var} 
      Comparison of the FONLL-B  and of the 5FS NNLO results for two different
      values of the matching scale $\mu_b=m_b$ (same as in
      Figs.~\ref{fig:muR_var}-\ref{fig:muF_var}) to each other and to the
      4FS NLO. Results are shown as a function of the renormalisation scale
      for fixed factorisation scale (top) or as a function of the
      factorisation scale for fixed renormalisation scale (bottom), in each
      case with the fixed scale chosen as
      $\mu=\frac{(m_Z+2m_b)}{3}$. For the case $\mu_b=2m_b$ only the
      upper edge of the uncertainty band is shown.}
  \end{center}
\end{figure}

Our results are summarised in
Figs.~\ref{fig:muR_var}-\ref{fig:m_mu_var}, where
matched results in the FONLL-A and FONLL-B scheme are compared to each
other and to the 4FS and 5FS scheme computations. Here and in the
sequel all results are given for LHC at 13~TeV.  Also, in
Table~\ref{tab:res}, results for the cross-section in the
three schemes with two different choices of central scale are
collected, with uncertainties obtained from
standard seven-point renormalization and factorization
scale variation.

In the three plots
we study respectively the renormalisation, factorisation, and matching
scale dependence of the results. 
In each
case, renormalisation and factorisation scales are fixed, and then
varied about, either a high value $\mu=m_Z$, or a low value
$\mu_{R,F}=\frac{m_Z+2m_b}{3}$. While the higher scale choice is standard in
inclusive $W$ and $Z$ production, the lower choice was advocated in
Refs.~\cite{Maltoni:2012pa,Lim:2016wjo} based on arguments 
that it is closer to the physical hard scale of the process, which
correponds to the average transverse momentum of the emitted partons, 
and leads
to faster perturbative convergence. With this scale choice clearly the
4FS and 5FS are generally in better agreement.

Because we extend the plots down to very low values of the
renormalization and factorization scales, for the preferred 
and most accurate  FONLL-B case,
we also provide an estimate
of the ambiguity of the scale-varied result, in order to be able to
assess whether and when the whole procedure becomes unreliable. This is done by
performing scale variation in
two different ways which differ by sub-leading terms. 
The two possibilities correspond to the
observation (see e.g.~\cite{Altarelli:2008aj})
that scale variation by a factor $k$ 
of a quantity $F(\mu)$ which is scale-independent up to NLO but has a
NNLO scale dependence can be performed by either letting
\begin{equation}\label{eq:scalup}
  F(\mu_0;k)=F(\mu_0+\ln k)- \ln k \frac{d}{d\ln\mu}
  F(\mu)\Big|_{\mu_0=\mu_0+\ln k},
\end{equation}
or
\begin{equation}\label{eq:scaldown}
  F(\mu_0;k)=F(\mu_0+\ln k)- \ln k \frac{d}{d\ln\mu} F(\mu)\Big|_{\mu=\mu_0},
\end{equation}
where the first term on the r.h.s. is computed up to NLO, while the
second term may be computed up to LO, and thus the two expressions
differ by NNLO terms (and similarly for higher orders). The two
options Eq.s~(\ref{eq:scalup}-\ref{eq:scaldown}) essentially correspond to
changing the sign of the scale-variation terms, i.e., they amount to
symmetrizing the scale variation: therefore, they are taken as the two
extremes of a band which provides an estimate of the 
uncertainty
on the scale uncertainty itself.
Finally, matching scale variation is performed by varying it between the
default $\mu_b=m_b$ and $\mu_b=2m_b$. When  $\mu_b=2m_b$, only results
for the choice Eq.~\ref{eq:scalup} of scale variation are shown, but
we have checked that the effect of varying the  matching scale when
renormalization and factorizaton scales
are varied according to  Eq.s~\ref{eq:scaldown} is similar, i.e., the
whole uncertainty band moves up and down when $\mu_b$ is varied
without changing shape significantly.


\begin{table}
  \begin{center}
    \begin{tabular}{lcccc}
      \label{tab:xs-res}
      & & $\scriptstyle \sigma^{4F}_{\text{NLO}}$ & $\scriptstyle \sigma^{5F}_{\text{NNLO}}$ & $\scriptstyle\sigma_{\text{FONLL-B}}$\\
      \hline
      &  &  &    &        \\
%      $\scriptstyle\mu = m_Z$ & & $\scriptstyle 935.36^{+13.9\%}_{-13.84\%}$~{\small pb} &
      $\scriptstyle\mu = m_Z$ & & $\scriptstyle 935.36^{+13.9\%}_{-13.8\%}$~{\small pb} &
      $\scriptstyle 1390.46^{+2.41\%}_{-3.07\%}$~{\small pb} &
      $\scriptstyle 1443.32^{+1.17\%}_{-3.14\%}$~{\small pb} \\
      &  &  &    &        \\
      \hline
      &  &  &    &        \\
      $\scriptstyle \mu = \frac{m_Z+2m_b}{3}$ & &
%      $\scriptstyle 1103.95^{+15.59\%}_{-13.9\%}$~{\small pb}  &
      $\scriptstyle 1103.95^{+15.6\%}_{-13.9\%}$~{\small pb}  &
      $\scriptstyle 1370.44^{+0.86\%}_{-5.89\%}$~{\small pb} &
      $\scriptstyle 1453.35^{+8.43\%}_{-2.71\%}$~{\small pb}\\
      &  &  &    &        \\
      \hline 
    \end{tabular}
    \caption{\label{tab:res} Summary of results for the bottom fusion
      cross-section.  Percentage error are
      obtained  as the envelope of a standard 7-point $\mu_R,\mu_F$
      variation arount the  central
      value $\mu_R=\mu_F=\mu$.}
  \end{center}
\end{table}

%% %% mu = mZ
%% \begin{align}
%% \sigma^{5F}_{NNLO}(m_Z)& = 1390.46^{+2.41\%}_{-3.07\%}~pb\nonumber \\
%% \sigma^{4F}_{NLO}(m_Z) & = 935.36^{+13.9\%}_{-13.84\%}~pb \nonumber \\
%% \sigma_{FONLL-B}(m_Z)    & = 1443.32^{+1.17\%}_{-3.14\%}~pb
%% \end{align}
%% %% Result 5F      :  1390.4624498721391  +  2.406406352180099  -  3.0682226972566236
%% %% Result 4F      :  935.3561343094833  +  13.90743647829309  -  13.838558632630804
%% %% Result FONLL-B :  1443.3170721973665  +  1.1721297793112746  -  3.1414516047307575

%% %% mu = (mZ+2mb)/3
%% \begin{align}
%% \sigma^{5F}_{NNLO}\left(\frac{m_Z+2m_b}{3}\right)& = 1390.46^{+0.86\%}_{-5.89\%}~pb\nonumber \\
%% \sigma^{4F}_{NLO}\left( \frac{m_Z+2m_b}{3}\right) & = 935.36^{+15.59\%}_{-13.9\%}~pb \nonumber \\
%% \sigma_{FONLL-B}\left(\frac{m_Z+2m_b}{3}\right)    & = 1453.35^{+8.43\%}_{-2.71\%}~pb
%% \end{align}

%% %% Result 5F      :  1370.4454233620424  +  0.8566373798124558  -  5.8913358077809255
%% %% Result 4F      :  1103.954607207936  +  15.578113201316135  -  13.904329790931609
%% %% Result FONLL-B :  1453.3488626612766  +  8.429187522436397  -  2.7133721442291447

We first describe and comment our
results, then discuss their interpretation, also in view of various
approximations which have been suggested in the literature. A first observation is that  comparison of
Figs.~\ref{fig:muR_var}-\ref{fig:muF_var} to the corresponding
plots for Higgs production in bottom fusion (Figs.~2-3 of
Ref.~\cite{Forte:2016sja}) show that they are qualitatively almost
indistinguishable: this is not unexpected given the similarity between
Higgs and $Z$ production which we already repeatedly emphasised. 

Coming now to these qualitative features we note that:
\begin{itemize}
\item The factorisation scale dependence is generally very slight,
  while the renormalisation scale dependence is, instead, stronger.
\item The scale dependence is quite large in the 4FS scheme, even at
  NLO though it is reduced in comparison to the LO case. It is much
  weaker in the 5FS and FONLL cases which all have similar and
  similarly weak scale dependence, except for very low values
  $\mu_R\sim \frac{m_Z}{10}$ where however the ambiguity on the scale
  uncertainty blows up. 
\item The perturbative expansion is  very
  unstable in the 4FS, with the LO and NLO results differing by a
  factor two or more. This instability is completely removed when the
  4FS is matched to the 5FS: indeed, the FONLL-A and FONLL-B are quite
  close to each other.
\item The 4FS and 5FS results are quite far from each other, with
  the 4FS NLO significantly
  closer to the 5FS than the LO. The FONLL results are in turn quite close to the 5FS.
\item The perturbative expansion is indeed more stable for a lower
  choice of factorisation and renormalisation scale. For very low
  scales $\mu\sim\frac{m_Z}{10}$ the 4FS and 5FS results become
  similar, but the scale dependence becomes very large: in fact, the
  width of the uncertainty band becomes as wide as the scale variation
  in comparison to central scale choice, meaning that the results
  become unreliable.
\item A change of matching scale has essentially the same effect on
  the 5FS and the FONLL results, and it has the effect of moving both
  towards the 45S, though by a moderate amount.
\end{itemize}

These qualitative features have a simple theoretical
interpretation. To this purpose, note that the cross-section for
this process contains collinear logarithms regulated by the heavy
quark mass, i.e. powers of $\ln\frac{\mu_Z}{m_b^2}$, one at each
perturbative order. These logs  
arise from a transverse
momentum integration, whose the upper limit is the maximum value of the
transverse momentum, i.e. the hard scale of the process, 
which is proportional to but not equal to $m_Z$, and the lower limit
is the physical production threshold, which is proportional to but not
equal to $m_b$. Of course, one can always rewrite the ensuing
logarithm as  $\ln\frac{\mu_Z^2}{m_b^2}$, plus constants (i.e. terms
which only depend on the dimensionless ratio $\tau=\frac{\mu_z^2}{s}$),
and mass corrections (i.e. terms suppressed by powers of
$\frac{\mu^2_b}{m_Z^2}$). 

In the 4FS the result is exact, so whatever is not included in the log is included in
the constants or in the mass corrections; on the other hand at
N$^{k}$LO only the first $k+1$ logs are included. In the 5FS the logs are
rewritten as
$\ln\frac{\mu_Z^2}{m_b^2}=\ln\frac{\mu_Z^2}{\mu_F^2}+\ln\frac{\mu_F^2}{\mu_b^2}+\ln\frac{\mu_b^2}{m_b^2}
$, where $\mu_F$ is the factorisation  scale and 
$\mu_b$ is the matching scale. The logs of the factorisation scale 
$\ln\frac{\mu_F^2}{\mu_b^2}$ are then resummed to all orders into
the evolution of the PDF, while
the logs of the hard scale $\ln\frac{\mu_Z^2}{\mu_F^2}$ are included to
finite order in the hard partonic cross-section and
logs the matching scale, $\ln\frac{\mu_b^2}{m_b^2}$, are 
included to finite order in the matching condition, which
expresses the initial $b$~PDF in terms of the gluon (they would be 
implicitly included in the initial PDF
if the $b$~PDF were independently parametrised).
When varying the factorisation scale, logs at the upper end
of the evolution are reshuffled between the resummed PDF and the
fixed-order but exact hard cross-section. When varying the matching
scale, logs at the bottom end 
of the evolution are reshuffled between the resummed PDF and the
fixed-order but exact hard matching condition.

Note that both the
hard coefficient and the
matching condition contains logs and constants, but not
mass-suppressed terms: so in the 5FS constants and logs of the
matching scale, as well as constants and logs in the hard
coefficient, 
are treated exactly but to fixed order, while logs of the
factorisation scale are  resummed to all
orders, but not constants. 
When the 5FS and the 4FS are matched into FONLL, also 
mass-suppressed terms, on top of constants
and logs of the matching scale, are treated
exactly.

The fact that the 4FS is perturbatively unstable while the 5FS is not
then is easily explained as a manifestation of the fact that the 4FS
contains large logs which are resummed in the 5FS. This is confirmed
by the fact that the large difference between the 4FS LO and NLO is of
the same order of the scale variation of the LO: indeed the scale
variation by construction captures the size of logarithmic
contribution. So the sizeable difference which persists between the 5FS
and NLO 4FS results is explained as being due to the higher order
(NNLO and beyond) logs
which are missing in the 4FS NLO, their size being quantitatively estimated
in ~\cite{Lim:2016wjo}. This is  confirmed by the observation
that the FONLL-A and FONLL-B include both the large log resummation,
and the full constants and mass suppressed terms, up to LO and NLO
respectively. The difference  between the FONLL-A and FONLL-B is thus
the size of the constant and mass-suppressed contributions to the
difference between the 4FS LO and NLO. This is seen to be much smaller
than the total difference between 4FS LO and NLO, which must therefore
be due to the log.

In order to further disentangle, within this small contribution, 
the constant from mass-suppressed term,
one would have to vary the hard scale, i.e. the $Z$ mass. This was
done in Ref.~\cite{Forte:2016sja} for Higgs production: variation of
the Higgs mass left the difference between FONLL-A and FONLL-B
essentially unchanged, thus showing that mass corrections are
negligible and the bulk of the difference between FONLL-A and FONLL-B
is due to a constant. Given the similarity between the two processes
we expect the same to be the case here. Given the small size of this
contribution the issue is largely academic anyway.

The qualitative form of the renormalisation scale dependence of the 4FS
result is also easy to understand: as the scale is decreased, the
value of $\alpha_s$ multiplying the large collinear log increases, and
both the LO and NLO predictions grow; this growth is only partly
reduced by the higher-order compensating term, at least down to
scales $\mu_R\sim 0.2\mu_Z$ where the ambiguity on the scale variation
itself becomes very large. The fact  that the 5FS (and FONLL) result
have almost no renormalisation scale dependence shows that this scale
dependence is coming from the $b$~quark term which is treated differently
between 4FS and 5FS.

The factorisation scale dependence is particularly intriguing. The
fact that this dependence is very slight in the 4FS is again consistent
with the observation that scale dependence is driven by the heavy
quark terms: in this scheme, in the absence of a $b$~PDF, the factorisation scale
dependence is  related to perturbative evolution of the light
quarks and gluons, which is moderate at NNLO. 
On the other hand, in the 5FS (and in FONLL) collinear logs are
resummed in the evolution  of the $b$-PDF up to $\mu_F$, and then
expanded out in the partonic cross-section from $\mu_F$ to the
physical hard scale of the process. We therefore expect the
factorisation scale dependence in this scheme to be approximately
stationary around this physical hard scale, very slight above it (where
$\alpha_s$ is small) and to only become significant when $\mu_F$ is
lower than the physical hard scale itself. This behaviour is
clearly seen in Fig.~\ref{fig:muF_var}, with the stationary point
close to the low scale advocated in Ref.~\cite{Maltoni:2012pa,Lim:2016wjo} 
that indeed this scale of the hard process, and it nicely explains the
very weak factorisation scale dependence also seen in the 5FS unless
$\mu_F\lesssim0.2m_Z$ or so.

Finally, the fact that when increasing the matching scale $\mu_b$ the
5FS and FONLL-B result decrease and get closer to the 4FS is
understood as a consequence of the fact that the resummed log becomes
smaller: as the matching scale is raised and gets closer to
the scale of the process the resummed logs become smaller, while the
fixed order log and constant included in the matching condition remain
the same. Since only these 
fixed order contributions are  included in the 4FS, as $\mu_b$ is
raised the 5FS gets closer to the 4FS.

We can finally discuss, in light of all this, the two related issues
of choosing the various scales, $\mu_F$, $\mu_R$ and $\mu_b$, and of
the validity of various approximations. As discussed, the scale
dependence of this process is driven by the collinear logs in the
$b$~ quark contribution, and thus the bulk of it comes from the
choice of argument in these logs. 

In a fully massive 4FS calculation,
these collinear logs are treated exactly, so the scale dependence
comes purely from the choice of argument in the strong coupling. It
then turns out that reducing the renormalisation scale increases the
4FS unresummed results up to the point where it agrees with the 5FS
resummed one. This is however accidental: the lack of resummation is made up
by artificially increasing $\alpha_s$, and indeed at low scale the
scale dependence of the 4FS result is not improved: if anything, it
increases. Hence, the 4FS appears to be a poor approximation to this
process and its improvement by lowering the renormalisation scale  is
unreliable.

In a 5FS calculation, instead, as mentioned, the exact upper and lower
limits of the transverse momentum integration are replaced by $\mu_F$
and $\mu_b$, respectively. As also mentioned, it has been
argued~\cite{Maltoni:2012pa,Lim:2016wjo} 
that the exact, kinematics-dependent upper limit of integration is
on average close to a scale $\frac{m_Z+2m_b}{3}\sim 0.35m_Z$. This is
borne out by our results: for all $\mu_F\gtrsim0.3m_Z$ the 
factorisation scale dependence of the 5FS result is flat, and with
this choice of $\mu_R$ the 5FS scale dependence is visibly
flatter. Given the smallness of mass
corrections, in practice a 5FS with low factorisation and
renormalisation scales appears to be a good approximation of the full
FONLL result.

On the other hand, it has been recently argued~\cite{Bertone:2017djs}
that a higher choice of matching scale may provide a better
approximation. Clearly, this is a process-dependent statement that
should be checked on a case-by-case basis: as discussed raising the
matching scale improves the accuracy of the starting, dynamically
generated PDF, as it matches it at a scale where perturbation theory
is more reliable, but it reduces the size of the logs which are
resummed. In the present case, the resummed logs are a large effect
and the constants a small correction, so raising the matching scale
does not appear to be advantageous: indeed, the renormalisation and
factorisation scale dependence is the same with $\mu_b=m_b$ or
$\mu_b=2m_b$, with no obvious improvement.

In fact, when raising $\mu_b$
the 5FS result decreases, and moves towards the low 4FS, but with no
improvement in perturbative stability of the latter. This is to be
contrasted to the case in which $\mu_R$ and $\mu_F$ are lowered, which
also brings the 4FS and the 5FS closer but now towards a high value,
and with a visible increase in perturbative stability. In fact, the
FONLL result shows that exact inclusion of the mass corrections (most
likely the constant) increases the pure 5FS, by a small
amount. On the contrary, raising the matching scale lowers it: this
means that the deterioration of the log resummation is a larger effect
than the improvement made by starting the PDF at a scale at which
perturbation theory is more reliable.
So a 5FS with
large $\mu_b$ does not appear to be a better approximation in our case:
it is likely to be a worse approximation if  $\mu_b$ is raised by a
moderate amount,
and it definitely appears to be a poor approximation if $\mu_b$ is
raised up to the point at which the 5FS result reduces to the 4FS
one. On the other hand, a variation of $\mu_b$ by perhaps a factor
two, as shown in Fig.~\ref{fig:m_mu_var}, might well be a reasonable
estimate of the uncertainty due to the use of a fixed-order matching
condition and should be included in the theoretical uncertainty, as was
done in Refs.~\cite{Bonvini:2016fgf,deFlorian:2016spz}. This theoretical
uncertainty can only be removed by parametrising the $b$~PDF, in which
case it is  traded for a PDF uncertainty.

Having 
 determined the total cross-section for $Z$
production in bottom quark fusion at the highest available accuracy in
a matched FONLL scheme, we can use this result in
order to improve the total $Z$ production cross-section. First, we
recall that, as already mentioned, there are further contributions
involving $b$ quarks in the final state to the $Z$ production
cross-section, but without initial-state bottom in the 5FS,
specifically at leading order the light-quark initated contribution of
Fig.~\ref{fig:fusvsfs}. Bottom mass effects in these contributions
could in priciple also be included in a
matched scheme. However, the corresponding 4FS contributions are only
known up to NLO, hence only an FONLL-A computation would be possible,
instead of the more accurate FONLL-B. Furthermore, also as already
mentioned, a matched computation including these contributions must be
performed at the level of the total $Z$ cross-section, rather than
that for  the $Zb\bar b$ cross-section, because these real emission
contribution 
are affected by infrared divergences which cancel against virtual
correction in which the $b$ quark circulates in loops but there are no
$b$ quarks in the final state. 

However, we can estimate the size of these contributions and the
impact of their FONLL improvement by computing
the leading-order contribution Fig.~\ref{fig:fusvsfs} by removing the
infrared divergence through  an
invariant mass cut $m_{b\bar b}\ge \sqrt{2} m_b$. We then get (with the
low scale choice and all other settings of the previous calculation) a
contribution $\sigma^{\rm light}_{\rm 5FS}(Zb\bar b)=146.1$~pb from the diagram
of Fig.~\ref{fig:fusvsfs} in the 5FS. The corresponding 4FS result is 
$\sigma^{\rm light}_{\rm 4FS}(Zb\bar b)=129.5$~pb, while the massless
limit of the 4FS result is $\sigma^{\rm light}_{\rm 0}(Zb\bar
b)=146.1$~pb. It follows that the effect of the FONLL-B improvement
over a pure 5FS computation of this term is at the level of less that
1\% of the total cross section of Table~\ref{tab:res}, to be compared
to the 5\%  impact of the FONLL-B improvement of the gluon fusion
contribution seen in  Table~\ref{tab:res}. But the very
definition of this contribution is subject to an ambiguity due to the
need to introduce a cutoff, which is of the order of the difference
between the whole contribution in the 5FS 
and its massless limit, namely, also of
the same size as the FONLL improvement. Therefore, its FONLL
improvement appears unnecessary.

\begin{figure}
  \begin{center}
    \includegraphics[width=0.45\textwidth]{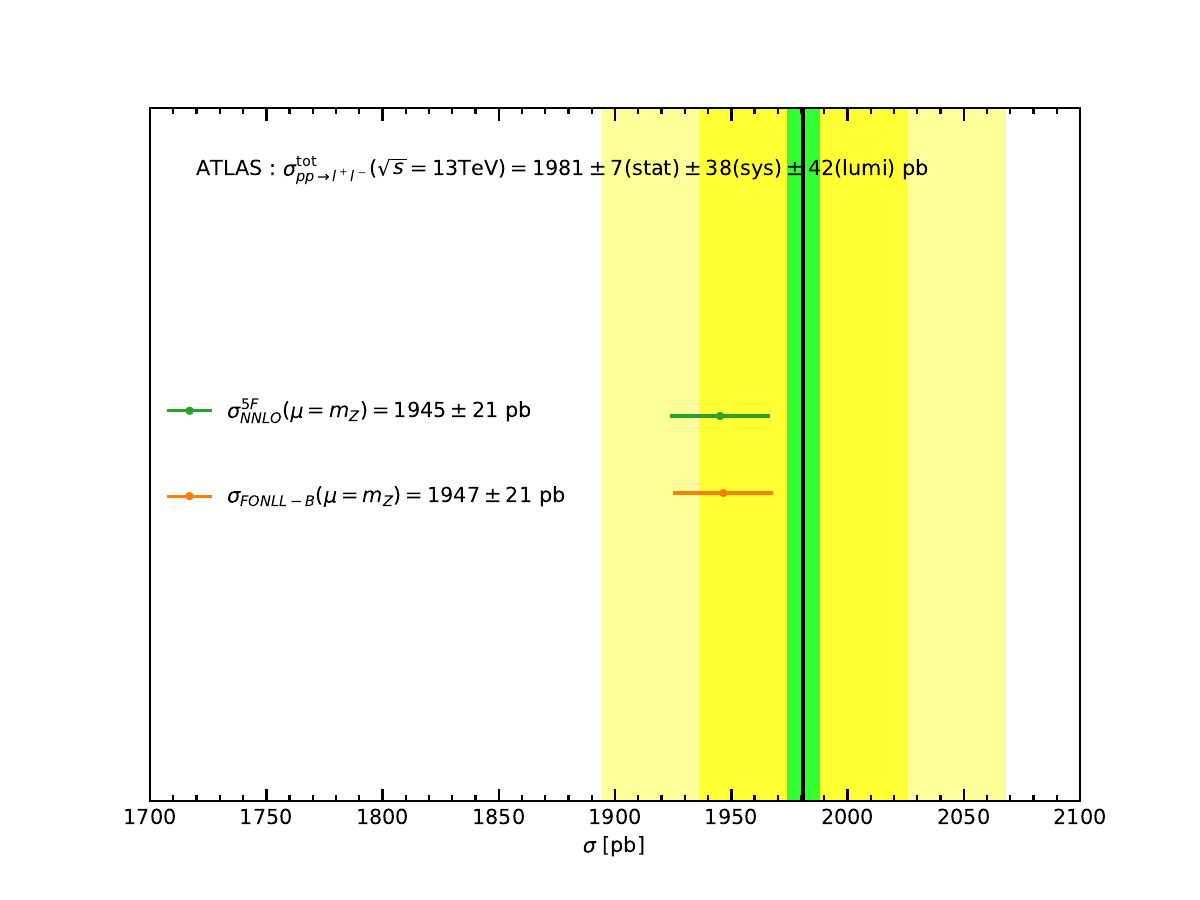}
    \includegraphics[width=0.45\textwidth]{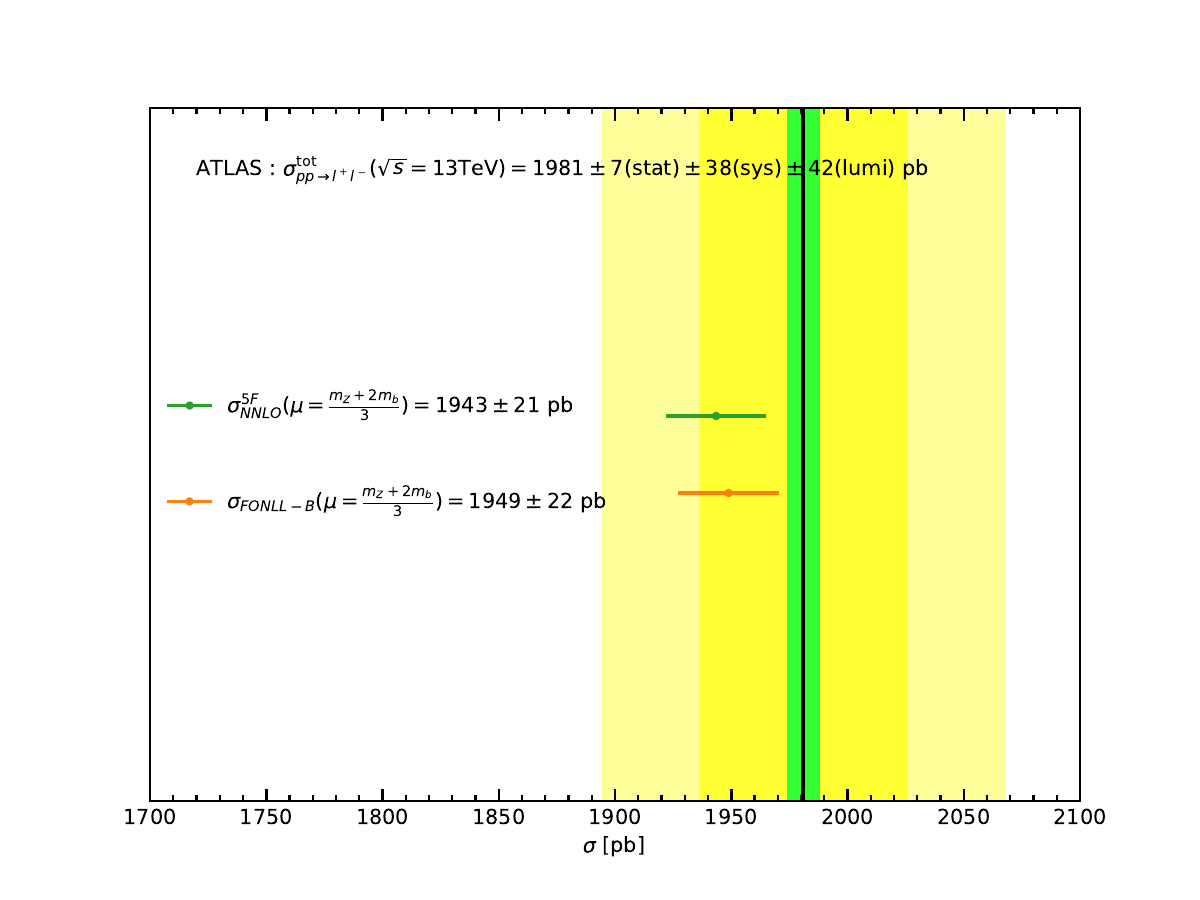} 
    \caption{\label{fig:ztotal}
The total cross-section for $Z$ production and decay into lepton
pairs, 
computed in a pure
five-flavour scheme, or with the bottom fusion contribution replaced
by its FONLL-B expression. Results are show for two different choices
of the renormalization and factorization scale  for the bottom-induced
contribution $\mu_R\mu_F=m_Z$ (left) or
$\mu_R\mu_F=\frac{m_Z+2m_b}{3}$ (right); in both cases the remaining
contributions to the total cross sections are evaluated with
$\mu_R\mu_F=m_Z$. The uncertainty band is obtained from seven-point
scale variation.
 Results are compared to the ATLAS measurement of
Ref.~\cite{Aad:2016naf}. }
  \end{center}
\end{figure}
We can thus consider the total cross-section. We compute this 
at NNLO in
the 5FS, using the same code and settings discussed above, and then
improve it by  subtracting
the bottom-initiated contribution to it, namely, the cross-section for
$Z$ production in bottom fusion, and  
replacing it with its FONLL expression as defined and discussed
above. Because both the total and the bottom-fusion cross-section are
separately collinear safe, this leads to a consistent result.
Results are shown in Fig.~\ref{fig:ztotal} for the rate into
lepton pairs, obtained multiplying by the branching ratio B(Z$\to$ll)$=3.3658~10^{-2}\pm 0.0023~10^{-2}$~\cite{Tanabashi:2018oca}. The uncertainty shown is
obtained from standard seven-point scale variation, with the central
result given as the mid-point of the band.  The bottom fusion
contribution is computed with either of the choices of scale that we
used, with the total cross-section determined with $\mu_R=\mu_F=m_Z$. 
At the level of total cross-section, the effect of the
bottom mass is very minor, at the permille level, much smaller
than the NNLO scale uncertainty. This justifies neglecting the further
FONLL improvement of the light-quark induced bottom production
copntribution, which at the level of total cross-section would be
possible, but, as we have seen above, would have a yet much smaller
impact. 

In summary, we have determined the total cross-section for $Z$
production in bottom quark fusion at the highest available accuracy in
a matched FONLL scheme,  and we have used our results as a test case for the
discussion of issues of scale dependence and heavy quark treatment, by
generalising our previous results for Higgs production, and studying
not only renormalisation and factorisation scale, but also matching
scale dependence. We have finally assessed the impact of the FONLL
improvement both on the bottom fusion and total $Z$ production
cross-section. 

Our main phenomenological conclusion is that,
similarly to the case of Higgs production,  mass effects on the bottom
fusion cross-section are small,
but non-negligible in comparison to the high experimental accuracy to
which this process can be measured. However, their impact on the total $Z$
cross-section is quite small, given that the bottom fusiom contribution
is only a small fraction of the total.
For bottom fusion, the contribution due to the
resummation of collinear logs of the heavy quark is sizeable, thereby
making a five-flavour scheme in which the $b$~quark is endowed with a PDF
a better approximation to the full FONLL result than the fixed-order
4FS calculation with massive b, which falls short of the full
prediction and displays large scale uncertainties. 
A low choice of
renormalisation and factorisation scale reduces the scale dependence
of both the full FONLL and pure 5FS result and is likely to improve
their accuracy, though in practice this makes little difference as the scale
dependence of both these results is very slight. However, it
does suggest that the hard physical scale for this process is
lower than the final-state mass, as previously advocated.

All in all, our results support the conclusion that, when dealing with
processes involving heavy quarks, a fully matched
treatment of heavy quarks with a proper inclusion of mass effects is
necessary for LHC phenomenology at the percent level, either through
its direct use, or as a guide to construct efficient and accurate
approximations. 

\bigskip
\bigskip
\begin{center}
  \rule{5cm}{.1pt}
\end{center}
\bigskip
\bigskip

A public implementation of our  NNLL+NLO FONLL-B matched computation
will be added to our code for Higgs
production~\cite{Forte:2016sja}, publicly available from 
\begin{center}
  \url{http://bbhfonll.hepforge.org/}.
\end{center}

\section*{Acknowledgments}
We thank Fabio Maltoni for providing the code for the calculation of the 5F NNLO
cross section and  Valerio Bertone for helping provided in generating
a PDF set with modified threshold  using the APFEL code. 
D.N. is supported by the French Agence Nationale de la Recherche, under
grant ANR-15-CE31-0016. S.F. is supported by the European Research Council
under the European Union’s Horizon 2020 research and innovation
Programme (grant agreement  n◦ 740006). M.U. is partially supported by the 
STFC grant ST/L000385/1 and by the Royal Society grants RGF/EA/180148
 and DH150088.
%%%%%%%%
\begin{appendix}
  \section{FONLL expressions with $\mu_b$ different from $m_b$}
  \numberwithin{equation}{section}
  \setcounter{equation}{0}
  We give for completeness the FONLL expressions by using $m_b$ different from $\mu_b$.
  Note that the only difference with respect to the formulae presented in~\cite{Forte:2016sja},
  is in the logarithm obtained from the expansion of the $b$~PDF, where $L=\log(Q^2/m_b^2)$,
  becomes $L=\log(Q^2/\mu_b^2)$.

  With this modification in place we get, for the 4F scheme, $B$ coefficients:
  \begin{align}
    B_{gg}^{(2)}\left(y,\frac{Q^2}{m^2_b}\right)        & = \hat{\sigma}_{gg}^{(2)}\left(y,\frac{Q^2}{m_b^2}\right) \\
    B_{q\bar{q}}^{(2)}\left(y,\frac{Q^2}{m^2_b}\right) & = \hat{\sigma}_{q\bar{q}}^{(2)}\left(y,\frac{Q^2}{m_b^2}\right)
  \end{align}
  while at $\order{\alpha_s^3}$ the redefinition of $\alpha_s$ contributes:
  \begin{align}
    B_{gg}^{(3)}\left(y,\frac{Q^2}{m^2_b},\frac{\mu_R^2}{\mu_b^2},\frac{\mu_F^2}{\mu_b^2}\right) & = \hat{\sigma}_{gg}^{(3)}\left(y,\frac{Q^2}{m_b^2}\right) - \frac{2 T_R}{3\pi} \ln{\frac{\mu_R^2}{\mu_F^2}}\hat{\sigma}_{gg}^{(2)}\left(y,\frac{Q^2}{m_b^2}\right)\\
    B_{q\bar{q}}^{(3)}\left(y,\frac{Q^2}{m^2_b},\frac{\mu_R^2}{\mu_b^2},\frac{\mu_F^2}{\mu_b^2}\right) & = \hat{\sigma}_{q\bar{q}}^{(3)}\left(y,\frac{Q^2}{m_b^2}\right)- \frac{2 T_R}{3\pi} \ln{\frac{\mu_R^2}{\mu_b^2}}\hat{\sigma}_{q\bar{q}}^{(2)}\left(y,\frac{Q^2}{m_b^2}\right) \\
    B_{gq}^{(3)}\left(y,\frac{Q^2}{m^2_b}\right)      & = \hat{\sigma}_{gq}^{(3)}\left(y,\frac{Q^2}{m_b^2}\right) \\
    B_{qg}^{(3)}\left(y,\frac{Q^2}{m^2_b}\right)      & = \hat{\sigma}_{qg}^{(3)}\left(y,\frac{Q^2}{m_b^2}\right).
  \end{align}

  The massless limit of the 4F scheme coefficients, $B^{(0)}$, are, in this case, given by
  \begin{equation}
    \label{eq:massless_lim}
    \sigma^{(4),(0)}\left(\alpha_s(Q^2),L\right)=\tau_H\int_{\tau_H}^{1} \frac{dx}{x}\int_{\frac{\tau_H}{x}}^{1} \frac{dy}
          {y^2}\sum_{ij=q,g}f_{i}(x,Q^2)f_j\left(\frac{\tau_H}{x y},Q^2\right)B_{ij}^{(0)}\left(y,L,\alpha_s(Q^2)\right),
  \end{equation}
  with
  \begin{equation}
    B_{ij}^{(0)}\left(y,L,\alpha_s(Q^2)\right) = \sum_{p=2}^N\left(\alpha_s(Q^2)\right)^pB_{ij}^{(0),(p)}\left(y,L\right),
  \end{equation}
  and
  \begin{align}
    B_{gg}^{(0)(2)} (y,L) & = y\int_y^1\frac{dz}{z}\left[2\mathcal{A}_{gb}^{(1)}\left(z,L\right)\mathcal{A}_{gb}^{(1)}\left(\frac{y}{z},L\right) + 4\mathcal{A}_{gb}^{(1)}\left(\frac{y}{z},L\right)\frac{1}{z}\hat{\sigma}_{gb}^{(1)}(z)\right] + \hat{\sigma}_{gg}^{(2)}(y), \\
    B_{q\bar{q}}^{(0)(2)} (y,L) &= \hat{\sigma}_{q\bar{q}}^{(2)}(y);
  \end{align}
  while the new contributions to  $\order{\alpha_s^3}$ are
  \begin{align}\label{eq:subtrexp}
    B_{gg}^{(0)(3)} (y,L) & = y\int_y^1\frac{dz}{z}\left[4\mathcal{A}_{gb}^{(2)}\left(z,L\right)\mathcal{A}_{gb}^{(1)}\left(\frac{y}{z},L\right) + 2\mathcal{A}_{gb}^{(1)}\left(z,L\right)\int_z^1\frac{dw}{w}\mathcal{A}_{gb}^{(2)}\left(\frac{y}{z},L\right)\frac{1}{w}\hat{\sigma}_{b\bar{b}}^{(1)}(z) \right.\nonumber \\
      & \left.\phantom{asdfdy\int_y^1\frac{dz}{z}4\mathcal{A}_{gb}^{(2)}\left(z,L\right)}+ 4\mathcal{A}_{gb}^{(2)}\left(\frac{y}{z},L\right)\hat{\sigma}_{gb}^{(1)}(z) + 4\mathcal{A}_{gb}^{(1)}\left(\frac{y}{z},L\right)\hat{\sigma}_{gb}^{(2)}(z)\right], \\
    B_{gq}^{(0)(3)} (y,L)  & =  y\int_y^1\frac{dz}{z}\left[2\mathcal{A}_{\Sigma b}^{(2)}\left(z,L\right)\mathcal{A}_{gb}^{(1)}\left(\frac{y}{z},L\right) + 2\mathcal{A}_{\Sigma b}^{(2)}\left(\frac{y}{z},L\right)\hat{\sigma}_{gb}^{(1)}(z) \right.\nonumber \\
      & \left.\phantom{asdfdy\int_y^1\frac{dz}{z}  4 \mathcal{A}_{gb}^{(2)}\left(z,L\right)\mathcal{A}_{gb}^{(1)}\left(\frac{y}{z},L\right)\mathcal{A}_{gb}^{(1)}\left(\frac{y}{z},L\right)}+ 2\mathcal{A}_{gb}^{(1)}\left(\frac{y}{z},L\right)\hat{\sigma}_{qb}^{(2)}(z)\right],
  \end{align}
  which completes our result in the case in which $\mu_b\neq m_b$.
\end{appendix}
%%%%%%%%%%%%%%%%
\renewcommand{\em}{}
\bibliographystyle{UTPstyle}
\bibliography{bbz_fonll}
\end{document}